\documentclass[letterpaper, 10 pt, conference]{Tim EMBC Paper/ieeeconf}  

\IEEEoverridecommandlockouts                              

\usepackage{cite}
\usepackage{graphicx}
\graphicspath{ {./Figures/} }
\usepackage{dblfloatfix}
\usepackage{multirow}
\usepackage{wasysym}

\title{\LARGE \bf
Comparison of Sub-Scalp EEG and Endovascular Stent-Electrode Array for Visual Evoked Potential Brain-Computer Interface
}

\author{Timothy B. Mahoney, Po-Chen Liu, David B. Grayden,~\IEEEmembership{Senior Member,~IEEE}, and Sam E. John,~\IEEEmembership{Member,~IEEE}%
\thanks{T.B. Mahoney, P.-C. Liu, D.B. Grayden, and S.E. John are with Dept of Biomedical Engineering, University of Melbourne, Victoria 3010, Australia. ({\tt tbmahoney@student.unimelb.edu.au})}
\thanks{T.B. Mahoney is supported by the Elizabeth and Vernon Puzey Scholarship.}
\thanks{D.B. Grayden is with Graeme Clark Institute, University of Melbourne, Victoria 3010, Australia.}
}

\begin{document}

\maketitle
\thispagestyle{empty}
\pagestyle{empty}

\begin{abstract}
Brain-computer interfaces (BCI) have the potential to improve the quality of life for persons with paralysis. Sub-scalp EEG provides an alternative BCI signal acquisition method that compromises between the limitations of traditional EEG systems and the risks associated with intracranial electrodes, and has shown promise in long-term seizure monitoring. However, sub-scalp EEG has not yet been assessed for suitability in BCI applications. This study presents a preliminary comparison of visual evoked potentials (VEPs) recorded using sub-scalp and endovascular stent electrodes in a sheep. Sub-scalp electrodes recorded comparable VEP amplitude, signal-to-noise ratio and bandwidth to the stent electrodes. 

\indent \textit{Clinical relevance}—This is the first study to report a comparision between sub-scalp and stent electrode array signals. The use of sub-scalp EEG electrodes may aid in the long-term use of brain-computer interfaces.
\end{abstract}

\section{INTRODUCTION}

Brain-computer interface (BCI) technology has the potential to improve the quality of life of persons with paralysis by enabling users to control a wheelchair, prosthesis, computer or other device using only their thoughts. However, despite considerable research efforts, BCIs are rarely seen outside lab environments. Electroencaphelography (EEG)-based BCIs require the user to don a headset fitted with numerous electrodes. Potential BCI users have stated that these headsets are ill-suited for daily use due to their cumbersome donning and doffing requirements, size, and appearance \cite{RN68, RN67, RN70, RN83, RN113, RN112}. More invasive approaches that implant electrodes on (electrocorticography or ECoG) or in (penetrating) the brain present significant risk to the patient and cause scaring that leads to reduced performance over time \cite{RN55, RN54, RN39, RN38, RN49, RN51}. 

Sub-scalp EEG provides an alternative BCI signal acquisition method that compromises between the limitations of traditional EEG systems and the risks of intracranial electrodes. As a device residing between the scalp and the skull, a sub-scalp implant has potential to provide a long-term, `set-and-forget' system that does not require craniotomy. A sub-scalp EEG BCI would eliminate the need for daily donning and doffing procedures, and would also include considerably improved comfort and aesthetic appeal to the user. Several sub-scalp EEG devices are currently at various stages of development around the world, mostly as chronic seizure monitoring solutions, and are well tolerated in the body for chronic periods, some of which having been implanted for over 24 months \cite{RN76, RN78, RN57}. In comparison to surface EEG systems, sub-scalp EEG has demonstrated higher bandwidth \cite{RN99} and improved signal quality \cite{RN59, RN63, RN78}, even moreso with \textit{peg} type electrodes that are embedded within the skull \cite{RN66}. 

A recent development in the field of BCI is the endovascular stent-electrode array, which accesses intracranial regions of the brain via the major blood vessels traversing the tissue \cite{RN104, Oxley102}. While the device is currently showing promise in clinical trials, it is not without limitations. The stent undergoes endothelialization and hence cannot be removed should there be a need \cite{7591716}. Additionally, recording spatial capabilities are limited to brain regions proximal to major blood vessels. In comparison, sub-scalp electrodes are removable and, while not reaching an equivalent depth, are more spatially versatile. 

To our knowledge, there has been no investigation into the feasibility of sub-scalp EEG for detection of BCI control signals. As such, this paper presents sub-scalp EEG data collected during a visual evoked potential (VEP) experiment. VEPs are often used for BCI control. Should the user be paralysed, yet maintain the use of their eyes, VEPs provide means for a BCI to determine where the subject is visually attending. Thus, if a BCI presents the user with various command options, each with their own stimulus pattern, the BCI can determine the subject's intent by mapping their VEP with the pattern they are attending to. Confirming that VEPs are detectable with sub-scalp EEG is a valuable first step toward providing a minimally-invasive BCI device for chronic use in the home. This paper presents results of a preliminary investigation into VEP detection using sub-scalp electrodes implanted in a sheep. 

Sheep models have been used previously for similar BCI studies as their anatomy is closer to human scale than smaller animals, such as monkeys \cite{8512385, RN105, RN66}. For comparison in this study, sub-scalp EEG was recorded in parallel with an endovascular stent-electrode array. The potential for sub-scalp EEG for BCI applications was assessed by evaluating the time-averaged VEP, signal-to-noise ratio (SNR) and maximum bandwidth.


\section{METHODS}

The experiment was approved by the Animal Ethics Committee of the Florey Institute of Neuroscience and Mental Health, Melbourne, Australia (22010). Brain signals were gathered from a single corriedale sheep using a stent-electrode array, subdural array and sub-scalp electrodes.
\begin{figure}[h] %
      \centering
      \vspace{-1.5em}
      \includegraphics[scale=0.8]{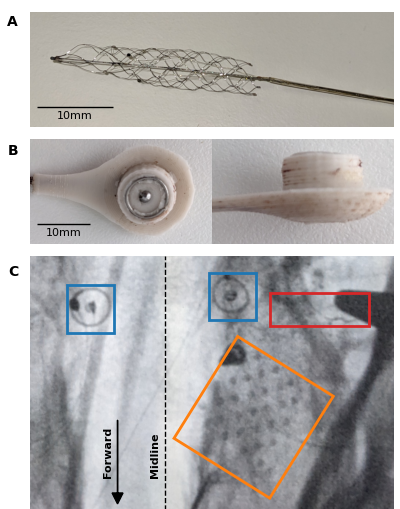}
      \caption{A. The endovascular stent-electrode array. B. One of the sub-scalp nodes, housing two electrodes in a Laplacian formation: an outer ring and centre ball. C. Top-down x-ray image of the implanted electrodes. The sub-scalp electrodes (blue squares) are positioned more medial and posterior than stent (red rectangle) and ECoG (orange square) electrode arrays.}
      \vspace{-1.8em}
      \label{Fig:electrodes}
   \end{figure}

\subsection{Electrodes}

The endovascular stent-electrode array (Figure \ref{Fig:electrodes}A) consists of four platinum disc electrodes ($\diameter$750~$\mu$m) mounted on a nitinol stent (Ev3, Inc., USA). The sub-scalp electrodes are stainless steel, and housed as pairs in a Laplacian (ring and ball) formation within a silicone structure, or node (Figure \ref{Fig:electrodes}B). All electrodes use a common reference.

\subsection{Surgical Procedure}

The sheep was anaesthetised during experimentation using a combination of isoflurane, propofol and fentanyl. Figure \ref{Fig:electrodes}C shows the electrode locations. The stent was inserted into the transverse sinus via the jugular vein. Two sub-scalp nodes were placed into the skull. The reference electrode was placed beneath the scalp, above the frontal lobe. An ECoG array was also placed over the brain, although the recordings are omitted from this report due to poor quality, likely due to an accute hemorrhage at the implantation site.

\subsection{VEP Experiment}
A full field flash (Grass Instrument Co., USA) was positioned approximately 20~cm from the animal's right eye and set to flash at 1.02~Hz. There were 5~min of stimulus and no-stimulus (baseline) activity recorded with both sub-scalp and stent electrodes simultaneously. Recordings were sampled at 512~Hz. Saline solution was applied to the eye at regular intervals to maintain eye health.

\subsection{Data Analysis}
A second-order Butterworth notch filter was used to reduce line noise and its harmonics. Stimulus data was segmented into epochs beginning 200~ms before stimulus and ending 1~s after stimulus. Epochs with a difference between maximum and minimum voltage greater than 150 $\mu$V were considered contaminated with artefact and were removed. An average over trials was computed to highlight the time-locked VEP. All analyses were performed in MATLAB 2020b (Mathworks Inc., USA).

\subsubsection{Signal-to-Noise Ratio}
Each epoch was split into a noise segment (200~ms pre-stimulus) and a VEP signal segment (400~ms post-stimulus). The SNR was determined as the ratio of the variance of the segments.  

\subsubsection{Frequency Response}
A fast Fourier transform (FFT) was performed on the baseline data to determine the bandwidth of each channel. A moving average filter was applied to the FFT with a width of 20~Hz to remove sharp fluctuations. The noise floor was estimated from the mean power in the 200-250~Hz band.

\subsubsection{Maximum Bandwidth}
Baseline data was segmented into 1~s epochs. Epochs with a difference between maximum and minimum voltage greater than 150 $\mu$V were considered contaminated with artefact and were removed. An FFT was performed on each epoch and averaged into 10~Hz bins. The first bin to drop below the noise floor was considered the maximum bandwidth measurable within the epoch.

\section{RESULTS}

The implantation and data collection of the sub-scalp and stent-electrode array was successful. Of the four channels recorded from each method, one sub-scalp and one stent channel were omitted from analysis due to poor signal quality, likely resulting from poor electrode contact. 

\begin{figure}[h] %
      \centering
      \includegraphics[scale=0.43]{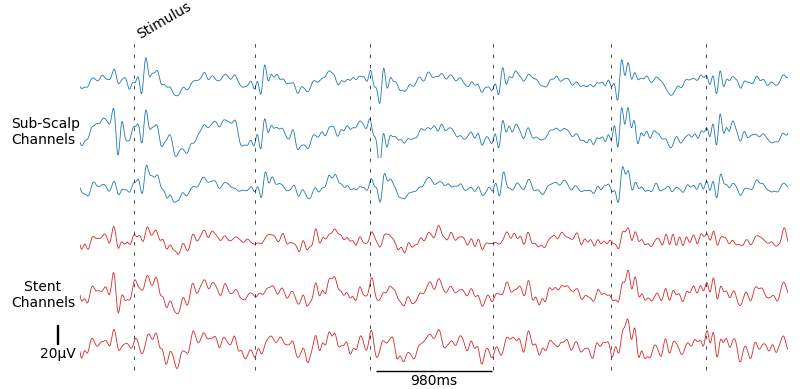}
      \caption{Data samples with stimulus times indicated by the dashed line for each channel of each recording method.}
      \label{Fig:trace}
   \end{figure}
   
\begin{figure}[h] %
      \centering
      \includegraphics[scale=0.42]{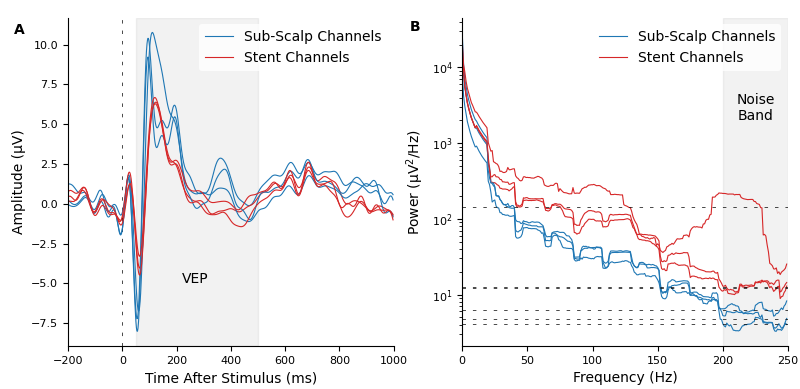}
      \vspace{-1.5em}
      \caption{A. The average VEP over trials. The shaded region outlines a clear VEP response following the stimulus (dashed line), recorded with three channels of both methods. B. The frequency spectrum of each channel captured during baseline recording (no-stimulus). Dashed lines indicate the mean power of the noise band for each channel (noise floor)}
      \label{Fig:AMPandFFT}
   \end{figure}

A clear VEP response was visible in each channel of the trace for both sub-scalp and stent electrodes (Figure \ref{Fig:trace}), though VEP peaks recorded from the stent were less clear. Despite this, averaging over trials revealed typical VEP responses (Figure \ref{Fig:AMPandFFT}A). Averaging over trials and channels revealed a 69\% increase in VEP amplitude with sub-scalp electrodes over stent electrodes (see Table \ref{table_1}).

\begin{figure}[thpb]
      \centering
      \includegraphics[scale=0.44]{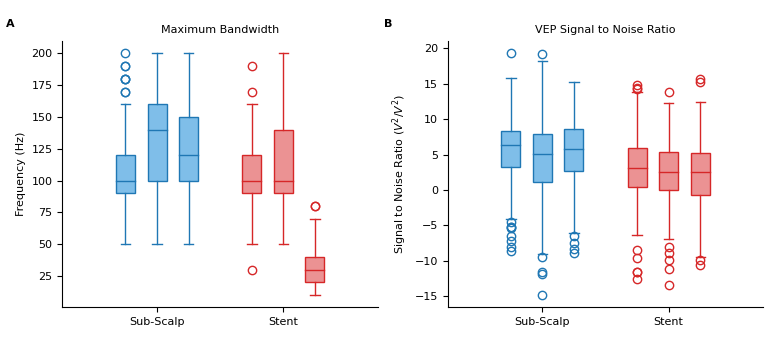}
      \vspace{-2em}
      \caption{A. The maximum bandwidth of each recording method. B. The VEP SNR of each recording method.}
      \vspace{-1.5em}
      \label{Fig:box}
   \end{figure}

Stent electrodes demonstrated higher power across the frequency spectrum, including the noise band (Figure \ref{Fig:AMPandFFT}B). Stent channel 3 showed a particularly high noise band power, resulting in a 39\% reduction in maximum bandwidth compared with other stent channels (Figure \ref{Fig:box}A and Table \ref{table_1}). On average, sub-scalp electrodes showed an increase in maximum bandwidth (123$\pm$38~Hz) over stent electrodes (87$\pm$43~Hz).

\begin{table}[h]
\centering
    \footnotesize
    \caption{Summary of Results}
    \label{table_1}
    \def\arraystretch{1.3}
    \begin{tabular}{p{0.1cm} p{0.7cm} p{2cm} p{1.8cm} p{1.7cm}}
        && VEP Amp. ($\mu$V) & SNR ($V^2$/$V^2$) & Max BW (Hz) \\
        \hline
        \multirow{4}{4em}{\rotatebox[origin=c]{90}{Sub-Scalp}} & CH1 &\hspace{7mm}17.5& \hspace{2mm}6.27$\pm8.07^1$&\hspace{3mm}105$\pm30^1$ \\
        & CH2 &\hspace{7mm}17.5 & \hspace{2mm}5.68$\pm$9.19&\hspace{3mm}139$\pm$38 \\
        & CH3 &\hspace{7mm}17.4 & \hspace{2mm}5.78$\pm$5.64&\hspace{3mm}128$\pm$36 \\
        & \textbf{Overall} &\hspace{7mm}\textbf{17.5}& \hspace{2mm}\textbf{5.91$\pm$7.75} &\hspace{3mm}\textbf{123$\pm$38}\\
        \hline
        \multirow{4}{4em}{\rotatebox[origin=c]{90}{Stent}} & CH1 &\hspace{7mm}10.5& \hspace{2mm}3.44$\pm$4.23&\hspace{3mm}104$\pm$27 \\
        & CH2 &\hspace{8.5mm}9.7& \hspace{2mm}2.82$\pm$3.15&\hspace{3mm}114$\pm$29 \\
        & CH3 &\hspace{7mm}10.8& \hspace{2mm}2.88$\pm$3.90&\hspace{4.5mm}34$\pm$16 \\
        & \textbf{Overall} &\hspace{7mm}\textbf{10.4}& \hspace{2mm}\textbf{3.05$\pm$3.79}&\hspace{4.5mm}\textbf{87$\pm$43} \\
        $^1$Mean$\pm$SD
    \end{tabular}
\end{table}

There was a significant difference (p$<$0.001) between the VEP SNR of both recording methods, with sub-scalp (5.91$\pm$7.75~Hz) showing an increase of 94\% over stent electrodes (3.05$\pm$3.79~Hz, Figure \ref{Fig:box}B). There was no significant inter-channel difference in SNR for each respective method (p$>$0.01, one-way analysis of variance (ANOVA)).

\section{DISCUSSION}

This study demonstrates that sub-scalp EEG electrodes are capable of detecting VEP responses. The VEP amplitude and SNR observed with sub-scalp electrodes was significantly more than simultaneous stent recordings. This result was unexpected, as bone tissue causes significant attenuation to brain signals recorded outside the skull. The improved signal quality in the sub-scalp electrodes was likely due to their positioning relative to the visual cortex. The sub-scalp electrodes were more medially positioned, placing them closer to the V1 visual cortex of the sheep, than the stent, resulting in higher amplitude VEPs. Additionally, the stent was not given time to adhere to the vessel wall via endothelialization and, as such, noise may be induced in the recording due to inadequate contact in an acute setting.

Stent electrodes recorded higher power across the spectrum than sub-scalp electrodes. This result is expected, as the stent electrodes do not see the attenuation effects of the skull. Despite this, sub-scalp electrodes are competitive with stent with regards to maximum bandwidth. This result is important as features in the high gamma band (80+ Hz) are useful in BCI applications, and are too attenuated to be isolated in surface EEG. This experiment demonstrated a maximum bandwidth over 100 Hz across each sub-scalp channel, as such, sub-scalp electrodes appear capable of accessing high gamma features.

Overall, sub-scalp EEG demonstrated comparable performance in recording VEPs and baseline EEG to the endovascular stent-electrode array. While only a preliminary study, these findings show promise that sub-scalp EEG would be a suitable signal acquisition method for a VEP-based BCI. Sub-scalp EEG has several advantages over stents. The device is easily implanted and removed, and can be positioned arbitrarily over the skull. Additionally, sub-scalp EEG addresses the limitations of surface EEG for BCI applications, namely its cumbersome setup requirements, discomfort to the user and poor aesthetic appeal.

\section{CONCLUSIONS}

This preliminary study presents a first step toward a potentially viable, chronic BCI solution that addresses the limitations of current invasive and non-invasive means of signal acquisition. There are many avenues for future studies in this space. More data is required to understand the benefits of certain types of sub-scalp electrodes, optimal electrode numbers and locations, hardware implementations and implantation procedures. With regards to peg electrodes, a longer implantation period may provide valuable insights into the effects of bone healing on the electrode and recording quality. Moving into human models, future work in this space should endeavour to further investigate sub-scalp EEG for BCI applications by evaluating performance with common paradigms (motor-imagery, steady-state visual evoked potentials, p300 etc.), and for use with particular applications, such as cursor control, mobility or communication. Research should also focus on gathering potential users' opinions of this promising method and engaging in user-centred design approaches to ensure BCI technology can finally leave the lab environment and enter the home.

\addtolength{\textheight}{0cm}   


\bibliographystyle{ieeetr}
\bibliography{refs.bib}

\begin{thebibliography}{10}

\bibitem{RN68}
A.~Kubler, E.~M. Holz, A.~Riccio, C.~Zickler, T.~Kaufmann, S.~C. Kleih,
  P.~Staiger-Salzer, L.~Desideri, E.~J. Hoogerwerf, and D.~Mattia, ``The
  user-centered design as novel perspective for evaluating the usability of
  bci-controlled applications,'' {\em PLoS One}, vol.~9, no.~12, p.~e112392,
  2014.
\newblock Research Support, Non-U.S. Gov't PLoS One. 2014 Dec 3;9(12):e112392.
  doi: 10.1371/journal.pone.0112392. eCollection 2014.

\bibitem{RN67}
F.~Miralles, E.~Vargiu, S.~Dauwalder, M.~Solà, G.~Müller-Putz, S.~C.
  Wriessnegger, A.~Pinegger, A.~Kübler, S.~Halder, I.~Käthner, S.~Martin,
  J.~Daly, E.~Armstrong, C.~Guger, C.~Hintermüller, and H.~Lowish, ``Brain
  computer interface on track to home,'' {\em The Scientific World Journal},
  vol.~2015, p.~623896, 2015.

\bibitem{RN70}
F.~Miralles, E.~Vargiu, X.~Rafael-Palou, M.~Solà, S.~Dauwalder, C.~Guger,
  C.~Hintermüller, A.~Espinosa, H.~Lowish, S.~Martin, E.~Armstrong, and
  J.~Daly, ``Brain–computer interfaces on track to home: Results of the
  evaluation at disabled end-users’ homes and lessons learnt,'' {\em
  Frontiers in ICT}, vol.~2, no.~25, 2015.

\bibitem{RN83}
B.~Peters, G.~Bieker, S.~M. Heckman, J.~E. Huggins, C.~Wolf, D.~Zeitlin, and
  M.~Fried-Oken, ``Brain-computer interface users speak up: The virtual users'
  forum at the 2013 international brain-computer interface meeting,'' {\em
  Archives of Physical Medicine and Rehabilitation}, vol.~96, no.~3,
  Supplement, pp.~S33--S37, 2015.

\bibitem{RN113}
J.~E. Huggins, P.~A. Wren, and K.~L. Gruis, ``What would brain-computer
  interface users want? opinions and priorities of potential users with
  amyotrophic lateral sclerosis,'' {\em Amyotroph Lateral Scler}, vol.~12,
  no.~5, pp.~318--24, 2011.
\newblock 1471-180x Huggins, Jane E Wren, Patricia A Gruis, Kirsten L
  Comparative Study Journal Article England 2011/05/04 Amyotroph Lateral Scler.
  2011 Sep;12(5):318-24. doi: 10.3109/17482968.2011.572978. Epub 2011 May 2.

\bibitem{RN112}
J.~E. Huggins, A.~A. Moinuddin, A.~E. Chiodo, and P.~A. Wren, ``What would
  brain-computer interface users want: opinions and priorities of potential
  users with spinal cord injury,'' {\em Arch Phys Med Rehabil}, vol.~96, no.~3
  Suppl, pp.~S38--45.e1--5, 2015.
\newblock 1532-821x Journal Article Research Support, U.S. Gov't, Non-P.H.S.
  United States 2015/02/28 Arch Phys Med Rehabil. 2015 Mar;96(3
  Suppl):S38-45.e1-5. doi: 10.1016/j.apmr.2014.05.028.

\bibitem{RN55}
B.~D. Winslow, M.~B. Christensen, W.-K. Yang, F.~Solzbacher, and P.~A. Tresco,
  ``A comparison of the tissue response to chronically implanted
  parylene-c-coated and uncoated planar silicon microelectrode arrays in rat
  cortex,'' {\em Biomaterials}, vol.~31, no.~35, pp.~9163--9172, 2010.

\bibitem{RN54}
B.~D. Winslow and P.~A. Tresco, ``Quantitative analysis of the tissue response
  to chronically implanted microwire electrodes in rat cortex,'' {\em
  Biomaterials}, vol.~31, no.~7, pp.~1558--1567, 2010.

\bibitem{RN39}
J.~D. Rolston, D.~Ouyang, D.~J. Englot, D.~D. Wang, and E.~F. Chang, ``National
  trends and complication rates for invasive extraoperative
  electrocorticography in the usa,'' {\em Journal of clinical neuroscience :
  official journal of the Neurosurgical Society of Australasia}, vol.~22,
  no.~5, pp.~823--827, 2015.
\newblock 25669117[pmid] PMC5501272[pmcid] S0967-5868(14)00717-6[PII].

\bibitem{RN38}
J.~D. Rolston, D.~J. Englot, S.~Cornes, and E.~F. Chang, ``Major and minor
  complications in extraoperative electrocorticography: A review of a national
  database,'' {\em Epilepsy research}, vol.~122, pp.~26--29, 2016.
\newblock 26921853[pmid] PMC5274526[pmcid] S0920-1211(16)30014-6[PII].

\bibitem{RN49}
Y.~Nagahama, A.~J. Schmitt, D.~Nakagawa, A.~S. Vesole, J.~Kamm, C.~K. Kovach,
  D.~Hasan, M.~Granner, B.~J. Dlouhy, M.~A. Howard, and H.~Kawasaki,
  ``Intracranial eeg for seizure focus localization: evolving techniques,
  outcomes, complications, and utility of combining surface and depth
  electrodes,'' {\em Journal of Neurosurgery JNS}, vol.~130, no.~4,
  pp.~1180--1192, 2019.

\bibitem{RN51}
D.~Taussig, G.~Dorfmüller, M.~Fohlen, C.~Jalin, C.~Bulteau,
  S.~Ferrand-Sorbets, M.~Chipaux, and O.~Delalande, ``Invasive explorations in
  children younger than 3years,'' {\em Seizure}, vol.~21, no.~8, pp.~631--638,
  2012.

\bibitem{RN76}
S.~Weisdorf, J.~Duun-Henriksen, M.~J. Kjeldsen, F.~R. Poulsen, S.~W. Gangstad,
  and T.~W. Kjær, ``Ultra-long-term subcutaneous home monitoring of
  epilepsy—490 days of eeg from nine patients,'' {\em Epilepsia}, vol.~60,
  no.~11, pp.~2204--2214, 2019.
\newblock https://doi.org/10.1111/epi.16360.

\bibitem{RN78}
R.~E. Stirling, M.~I. Maturana, P.~J. Karoly, E.~S. Nurse, K.~McCutcheon, D.~B.
  Grayden, S.~G. Ringo, J.~M. Heasman, R.~J. Hoare, A.~Lai, W.~D'Souza,
  U.~Seneviratne, L.~Seiderer, K.~J. McLean, K.~J. Bulluss, M.~Murphy, B.~H.
  Brinkmann, M.~P. Richardson, D.~R. Freestone, and M.~J. Cook, ``Seizure
  forecasting using a novel sub-scalp ultra-long term eeg monitoring system,''
  {\em Frontiers in Neurology}, vol.~12, no.~1445, 2021.

\bibitem{RN57}
J.~Duun-Henriksen, M.~Baud, M.~P. Richardson, M.~Cook, G.~Kouvas, J.~M.
  Heasman, D.~Friedman, J.~Peltola, I.~C. Zibrandtsen, and T.~W. Kjaer, ``A new
  era in electroencephalographic monitoring? subscalp devices for
  ultra-long-term recordings,'' {\em EPILEPSIA}, 2020.

\bibitem{RN99}
J.~D. Olson, J.~D. Wander, L.~Johnson, D.~Sarma, K.~Weaver, E.~J. Novotny,
  J.~G. Ojemann, and F.~Darvas, ``Comparison of subdural and subgaleal
  recordings of cortical high-gamma activity in humans,'' {\em Clinical
  neurophysiology : official journal of the International Federation of
  Clinical Neurophysiology}, vol.~127, no.~1, pp.~277--284, 2016.
\newblock 25907415[pmid] PMC4600028[pmcid] S1388-2457(15)00230-8[PII].

\bibitem{RN59}
G.~B. Young, J.~R. Ives, M.~G. Chapman, and S.~M. Mirsattari, ``A comparison of
  subdermal wire electrodes with collodion-applied disk electrodes in long-term
  eeg recordings in icu,'' {\em Clinical Neurophysiology}, vol.~117, no.~6,
  pp.~1376--1379, 2006.

\bibitem{RN63}
J.~Duun-Henriksen, T.~W. Kjaer, D.~Looney, M.~D. Atkins, J.~A. Sørensen,
  M.~Rose, D.~P. Mandic, R.~E. Madsen, and C.~B. Juhl, ``Eeg signal quality of
  a subcutaneous recording system compared to standard surface electrodes,''
  {\em Journal of Sensors}, vol.~2015, p.~341208, 2015.

\bibitem{RN66}
Y.~B. Benovitski, A.~Lai, C.~C. McGowan, O.~Burns, V.~Maxim, D.~A.~X. Nayagam,
  R.~Millard, G.~D. Rathbone, M.~A. le~Chevoir, R.~A. Williams, D.~B. Grayden,
  C.~N. May, M.~Murphy, W.~J. D’Souza, M.~J. Cook, and C.~E. Williams, ``Ring
  and peg electrodes for minimally-invasive and long-term sub-scalp eeg
  recordings,'' {\em Epilepsy Research}, vol.~135, pp.~29--37, 2017.

\bibitem{RN104}
N.~L. Opie, N.~R. van~der Nagel, S.~E. John, K.~Vessey, G.~S. Rind, S.~M.
  Ronayne, E.~L. Fletcher, C.~N. May, O.~B. TJ, and T.~J. Oxley, ``Micro-ct and
  histological evaluation of an neural interface implanted within a blood
  vessel,'' {\em IEEE Trans Biomed Eng}, vol.~64, no.~4, pp.~928--934, 2017.
\newblock 1558-2531 IEEE Trans Biomed Eng. 2017 Apr;64(4):928-934. doi:
  10.1109/TBME.2016.2552226. Epub 2016 Jun 21.

\bibitem{Oxley102}
T.~J. Oxley, P.~E. Yoo, G.~S. Rind, S.~M. Ronayne, C.~M.~S. Lee, C.~Bird,
  V.~Hampshire, R.~P. Sharma, A.~Morokoff, D.~L. Williams, C.~MacIsaac, M.~E.
  Howard, L.~Irving, I.~Vrljic, C.~Williams, S.~E. John, F.~Weissenborn,
  M.~Dazenko, A.~H. Balabanski, D.~Friedenberg, A.~N. Burkitt, Y.~T. Wong,
  K.~J. Drummond, P.~Desmond, D.~Weber, T.~Denison, L.~R. Hochberg, S.~Mathers,
  T.~J. O{\textquoteright}Brien, C.~N. May, J.~Mocco, D.~B. Grayden, B.~C.~V.
  Campbell, P.~Mitchell, and N.~L. Opie, ``Motor neuroprosthesis implanted with
  neurointerventional surgery improves capacity for activities of daily living
  tasks in severe paralysis: first in-human experience,'' {\em Journal of
  NeuroInterventional Surgery}, vol.~13, no.~2, pp.~102--108, 2021.

\bibitem{7591716}
N.~L. Opie, G.~S. Rind, S.~E. John, S.~M. Ronayne, D.~B. Grayden, A.~N.
  Burkitt, C.~N. May, T.~J. O'Brien, and T.~J. Oxley, ``Feasibility of a
  chronic, minimally invasive endovascular neural interface,'' in {\em 2016
  38th Annual International Conference of the IEEE Engineering in Medicine and
  Biology Society (EMBC)}, pp.~4455--4458, 2016.

\bibitem{8512385}
N.~L. Opie, S.~E. John, G.~S. Rind, S.~M. Ronayne, C.~N. May, D.~B. Grayden,
  and T.~J. Oxley, ``Effect of implant duration, anatomical location and
  electrode orientation on bandwidth recorded with a chronically implanted
  endovascular stent-electrode array,'' in {\em 2018 40th Annual International
  Conference of the IEEE Engineering in Medicine and Biology Society (EMBC)},
  pp.~1074--1077, 2018.

\bibitem{RN105}
S.~E. John, N.~L. Opie, Y.~T. Wong, G.~S. Rind, S.~M. Ronayne, G.~Gerboni,
  S.~H. Bauquier, T.~J. O’Brien, C.~N. May, D.~B. Grayden, and T.~J. Oxley,
  ``Signal quality of simultaneously recorded endovascular, subdural and
  epidural signals are comparable,'' {\em Scientific Reports}, vol.~8, no.~1,
  p.~8427, 2018.

\end{thebibliography}
\end{document}